# Advanced cervical cancer classification: enhancing pap smear images with hybrid PMD filter-CLAHE

Ach Khozaimi[1,2], Isnani Darti[1], Syaiful Anam[1], Wuryansari Muharini Kusumawinahyu[1]
[1]Department of Mathematics, Faculty of Mathematics and Natural Sciences, Brawijaya University, Malang, Indonesia
[2]Department of Computer Science, Faculty of Engineering, Universitas Trunojoyo Madura, Madura, Indonesia

**ABSTRACT**

Cervical cancer remains a significant health problem, especially in developing countries. Early detection is critical for effective treatment. Convolutional neural networks (CNN) have shown promise in automated cervical cancer screening, but their performance depends on pap smear image quality. This study investigates the impact of various image preprocessing techniques on CNN performance for cervical cancer classification using the SIPaKMeD dataset. Three preprocessing techniques were evaluated: Perona-Malik diffusion (PMD) filter for noise reduction, contrast-limited adaptive histogram equalization (CLAHE) for image contrast enhancement, and the proposed hybrid PMD filter-CLAHE approach. The enhanced image datasets were evaluated on pretrained models, such as ResNet-34, ResNet-50, SqueezeNet-1.0, MobileNet-V2, EfficientNet-B0, EfficientNet-B1, DenseNet-121, and DenseNet-201. The results show that hybrid preprocessing PMD filter-CLAHE can improve the pap smear image quality and CNN architecture performance compared to the original images. The maximum metric improvements are 13.62% for accuracy, 10.04% for precision, 13.08% for recall, and 14.34% for F1-score. The proposed hybrid PMD filter-CLAHE technique offers a new perspective in improving cervical cancer classification performance using CNN architectures.

*Corresponding Author:*

Syaiful Anam
Department of Mathematics, Faculty of Mathematics and Natural Sciences, Brawijaya University
Malang, 65145, Indonesia
Email: syaiful@ub.ac.id

## 1. INTRODUCTION

Cervical cancer is the fourth most common cancer in women worldwide [1]. According to the global cancer observatory (GCO), cervical cancer was responsible for an estimated 570,000 new diagnoses and 311,000 fatalities worldwide in 2018 [2]. In developing nations, nearly 90% of the deaths are related to cervical cancer [3]. A commonly employed technique for early cancer detection is the pap smear test, which involves extracting cells from the cervix and analyzing them to identify precancerous or cancerous changes [4]. Despite its effectiveness, traditional pap smear analysis is labor-intensive and subject to human error, necessitating the development of automated and reliable diagnostic tools [5]. With the advent of advanced machine learning techniques, particularly convolutional neural networks (CNN), there is an opportunity to enhance the accuracy and efficiency of cervical cancer detection using automated image analysis [6]. CNN has revolutionized medical image classification because it can automatically extract and learn features from complex datasets [7]. However, the performance of CNN models is heavily influenced by image quality [8]. Most pap smear images exhibit low contrast and significant noise [9]. Noise reduction and contrast enhancement have been proposed to improve CNN performance [10]. Image preprocessing is a crucial step that can significantly affect the outcomes of deep learning models. Image contrast is required to produce an accurate model because classification features are usually based on nuclear and cytoplasmic characteristics [11].

Several studies have demonstrated that reducing image noise significantly enhances classification performance [12]. The Perona-Malik diffusion (PMD) filter removes noise, smoothens images, and maintains crucial edge details [13]. It applies a modified Gaussian function, assigning greater weights to central pixels and lower weights to those at the edges [14]. Research indicates that PMD filters play a key role in detecting and extracting malignant tumors in medical images [15] and have been shown to enhance deep learning performance in cervical cancer classification [16]. In addition, noise removal and contrast enhancement are essential for improving pap smear image quality. Contrast-limited adaptive histogram equalization (CLAHE), an improved version of adaptive histogram equalization (AHE), restricts excessive contrast enhancement to prevent artifacts [17]. It has significantly enhanced pap smear images and improved the classification accuracy of deep learning models such as visual geometry group-16 (VGG16), InceptionV3, and EfficientNet in cervical cancer diagnosis [18]. CLAHE has also been successful in improving image quality and boosting the performance of machine learning algorithms such as artificial neural networks (ANN) and K-nearest neighbors (KNN) in cervical cancer classification [19]. Additionally, research has shown that CLAHE enhances the ability of the you only look once (YOLO) algorithm to recognize road markings at night [20], improves CNN-based segmentation of lung cancer in computed tomography (CT) scan images [21], and enhances water-image classification [22]. Given their success, it is essential to investigate whether combining both techniques can improve classification performance.

This study explores the effects of a hybrid PMD filter-CLAHE on cervical cell image quality. While PMD filters are known for noise reduction and edge preservation, their impact on contrast enhancement is less studied. Similarly, CLAHE improves contrast, but its role in noise suppression and structural conservation in medical imaging is unclear. This study examines their combined effects to address these gaps. The PMD filter reduces noise while preserving edge details and important image structures [23]. By contrast, CLAHE enhances image contrast and improves the visibility of critical features for CNN-based classification [24]. Studies have shown that CLAHE-based entropy analysis is more effective than other methods in enhancing medical images [25]. To assess the effectiveness of these preprocessing techniques, we conducted experiments using the SIPaKMeD dataset, a widely used dataset for cervical cancer classification [26]. The study was evaluated using CNN architectures, including ResNet34, ResNet50, SqueezeNet-1.0, MobileNet-V2, EfficientNet-B0, EfficientNet-B1, DenseNet121, and DenseNet201, to measure their classification performance using different preprocessing methods. The key evaluation metrics included image quality assessment, accuracy, precision, recall, F1-score, and training time. Image quality was assessed using the contrast enhancement-based image quality (CEIQ) metric [27].

CNN architectures were selected based on their unique strengths in deep learning applications. ResNet introduces residual learning with skip connections, which effectively mitigates the vanishing gradient problem and enables the training of very deep networks [28]. SqueezeNet, designed for efficiency, uses fire modules to reduce the number of parameters, making it an ideal choice for tasks that require lightweight models with minimal memory usage while maintaining competitive accuracy [29]. MobileNet, optimized for mobile and embedded vision applications, employs depth-wise separable convolutions, significantly reducing the model size and computational requirements for efficient processing [30]. EfficientNet balances network depth, width, and resolution, achieving state-of-the-art accuracy with fewer parameters and floating-point operations per second (FLOPs) than other models [31]. Finally, DenseNet establishes dense connections between layers, maximizing information flow and gradient propagation, which enhances feature reuse and mitigates the vanishing gradient problem [32].

The experimental results show that each preprocessing technique offers unique advantages, with the most significant improvement observed using the hybrid PMD filter-CLAHE. These findings emphasize the importance of selecting and integrating preprocessing techniques to enhance medical image quality and improve diagnostic accuracy. PMD filtering is particularly effective for lightweight architectures, and CLAHE benefits deeper CNN architectures. Most importantly, the hybrid PMD filter-CLAHE approach consistently improves performance across most CNN architectures, balancing noise reduction and contrast enhancement to optimize feature extraction. This study provides valuable insights for refining image preprocessing strategies. This contributes to more precise and reliable cervical cancer detection using deep learning models. We found that image contrast enhancement and noise reduction correlate with feature extraction efficiency, as higher contrast leads to better-defined cell structures. The proposed method in this study tended to have an inordinately higher proportion of well-enhanced image regions as significant features in CNN activations, improving classification performance across different architectures.

## 2. METHOD
### 2.1. Material, methods, and simulation

Figure 1 illustrates the sequential steps of preprocessing pap smear images for cervical cancer classification using deep learning. The process begins with input data using the SIPaKMeD dataset. Next, the image preprocessing included a PMD filter and CLAHE. After preprocessing, the dataset was split into training (80%), validation (10%), and test (10%) data to ensure proper model training, fine-tuning, and evaluation. The enhanced images were evaluated using deep learning architectures, such as ResNet, EfficientNet, MobileNet, and DenseNet. The training set was employed to develop the model, whereas the validation set was utilized to optimize hyperparameters and mitigate overfitting issues. After training, the model was evaluated using test data. Performance metrics such as accuracy, recall, F1-score, and precision were calculated.

The SIPaKMeD dataset is a publicly available collection designed to support research on automated cervical cancer classification [26]. The SIPaKMeD dataset consisted of 4,049 high-resolution cervical cell images extracted from pap smear slides and manually annotated by cytopathology experts. These images were acquired using a 40× objective lens to ensure that detailed cellular structures were preserved. The dataset included isolated single-cell images, indicating that each image contained a single cervical cell rather than clusters of multiple cells. The images showed size, contrast, illumination, and staining intensity variations. The SIPaKMeD dataset was divided into five categories: superficial squamous epithelial cells, intermediate squamous epithelial cells, high-grade squamous intraepithelial lesion (HSIL) cells, columnar epithelial cells, and low-grade squamous intraepithelial lesion (LSIL) cells. Each category comprised approximately 800–850 images. SIPaKMeD is widely used in deep learning-based cervical cancer classification and is a benchmark for testing image preprocessing techniques, feature extraction methods, and CNN architectures. This dataset provides a realistic challenge for developing robust and generalizable cervical cancer screening models [33].

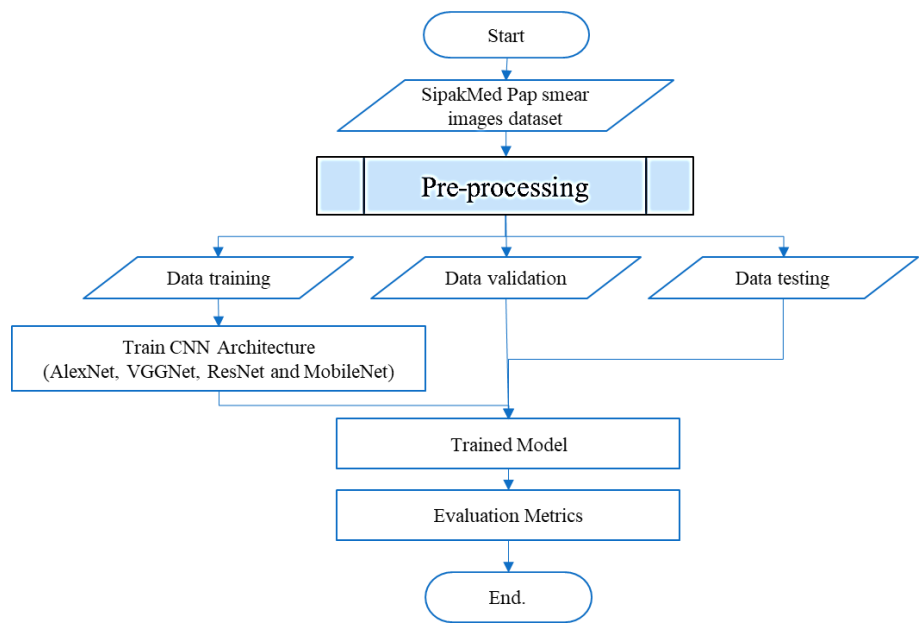

Figure 1. Method and simulation flowchart

Pap smear images often contain noise with a low contrast [9]. Preprocessing is crucial for enhancing the image quality for cervical cancer classification. This study explored the impact of different image preprocessing techniques on CNN performance, specifically focusing on the PMD filter, CLAHE, and a hybrid PMD filter-CLAHE approach. Figure 2 shows a flowchart depicting the preprocessing steps applied to the pap smear images from the SIPaKMeD dataset using a combination of the PMD filter and CLAHE techniques. Initially, the pap smear images were split into color spaces: red (R), green (G), and blue (B). A PMD filter and/or CLAHE were applied to each color space. This sequential application aims to enhance the contrast and denoise each color component. The R, G, and B components are enhanced using a PMD filter or CLAHE, resulting in improved and denoised R, G, and B components. After processing each color component, the contrast enhancement and denoised R-, G-, and B-components were combined to obtain the final enhanced pap smear images. The PMD filter helps reduce noise while preserving essential cell structures. It works by smoothing uniform areas while maintaining edges, ensuring that crucial details are not lost. PMD filter has three main parameters: iterations (N), diffusion coefficient (κ), and time step (λ). More iterations enhance noise removal but may blur the details. The diffusion coefficient controls edge preservation; low values keep edges sharp, whereas high values smooth the image. The time step affects stability and should be ≤0.25; smaller values filter gradually, while larger values speed up diffusion but risk over-smoothing [34]. In this study, we set N=20, κ=20, and λ=0.25. A detailed explanation of the PMD filter is provided in section 2.2. CLAHE enhances contrast by locally adjusting the brightness in small image regions, thereby preventing over-enhancement in bright or dark areas. CLAHE has two main parameters: clip limit and block size. The clip limit controls contrast and low values prevent over-enhancement, whereas high values boost contrast but may add noise. The block size defines the local regions. Small blocks enhance details, while large blocks give smoother results [34]. In this study, we set the clip limit to 2.0 and the block size to 8×8. A description of the CLAHE is provided in section 2.3. Processing each RGB channel separately ensures better noise removal and contrast adjustment while preserving the natural color balance of the cervical cell images [35].

　　　　This study conducted five simulations to evaluate the impact of different image preprocessing techniques on enhancing pap smear images. The first simulation tested ten pap smear images using three preprocessing techniques: a PMD filter, CLAHE, and a hybrid PMD filter-CLAHE. The goal was to assess how these techniques improve the image quality. The second experiment used the original pap smear images as the baseline. The Third simulation applied the PMD filter to assess its effectiveness in reducing noise in pap smear images. The fourth simulation used CLAHE to enhance the contrast of the pap smear images. The fifth simulation applied the hybrid PMD filter-CLAHE to test whether their combined are more effective in improving pap smear image quality. The first simulation was evaluated using CEIQ for image quality, and the other simulation was evaluated using CNN architectures. These evaluations helped determine the most effective preprocessing approach for optimizing CNN-based cervical cancer classification methods. Torch vision models were used in this study, including pretrained ResNet34, Resnet50, SqueezeNet-1.0, MobileNet-V2, EfficientNet-B0, EfficientNet-B1, DenseNet121, and DenseNet201. Adam optimizer was used in learning models with 0.0001 for learning rate. This study conducted training over 30 epochs, allowing sufficient iterations for the model to converge while mitigating the risk of overfitting. Combining the pretrained TorchVision models, the Adam optimizer, the learning rate, and the appropriate epoch count enabled CNN to train the models efficiently. The simulations used a Python Jupyter Notebook with Torch vision models on a Windows 11 operating system. TorchVision models offer a wide range of state-of-the-art CNN architectures [36]. The hardware configuration consisted of an AMD Ryzen 5 5500 processor complemented by 32 GB of system memory. A NVidia GeForce RTX 3060 graphics processing unit (GPU) with 12 GB of dedicated VRAM was used for accelerated CNN architecture computations.

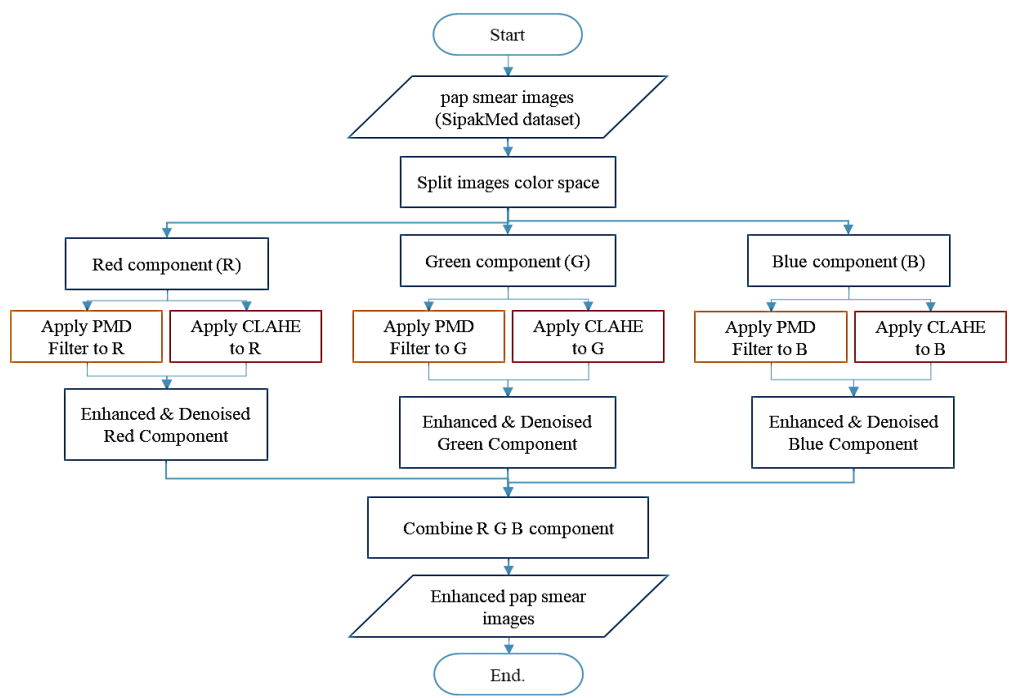

Figure 2. Pap smear image preprocessing using PMD filter, CLAHE, or hybrid PMD filter –CLAHE

## 2.2. PMD filter

The PMD filter operates on the principle of anisotropic diffusion, where the diffusion rate is adapted based on the gradient variations of the image. It facilitates diffusion primarily in areas with low gradients (uniform regions), while restricting it to high-gradient regions (edges). This controlled diffusion process is dictated by a partial differential equation (PDE), which regulates the flow of information within the image [37]. The following nonlinear diffusion equation describes the PMD Filter:

$$\frac{\partial I(x,y,t)}{\partial t} = \nabla \cdot (c(\| \nabla I \|)\nabla I), \quad (1)$$

$I(x, y, t)$ is the image intensity at the position $(x, y)$ and time $t$, $\nabla I$ is the gradient of the image, $c(\| \nabla I \|)$ is the diffusion coefficient, $\nabla \cdot$ denotes the divergence operator. The diffusion coefficient regulates the rate of diffusion, which is influenced by the magnitude of the gradient. $|\nabla I|$. The two commonly used diffusion coefficient functions are inverse quadratic and exponential functions. The exponential was performed in this study.

$$c(\| \nabla I \|) = e^{-\left(\frac{|\nabla I|}{K}\right)^2} \quad (2)$$

Where $K$ is a contrast that determines the sensitivity to edges. Small values of $K$ make the filter more sensitive to smaller edges, while larger values make it less sensitive. The PMD filter is a powerful tool for image enhancement, offering significant advantages over traditional linear smoothing techniques. Its ability to selectively smooth images while preserving edges makes it highly valuable in various applications [38].

## 2.3. CLAHE

CLAHE is utilized to enhance image contrast [39]. It operates by dividing the image into small regions. It applies histogram equalization (HE) to each region using (3) and controlling contrast enhancement by limiting amplification using (4). This method effectively prevents excessive contrast enhancement in uniform areas, and is particularly advantageous for image processing. CLAHE improves fine details while reducing artifacts. CLAHE is effective in improving the medical images quality [25].

$$K_o = round\left(\frac{C_i \cdot (2k-1)}{w.h}\right) \quad (3)$$

In (4) is used to obtain the new grey value of the histogram equalization result. $K_o$. $C_i$ is the cumulative distribution of the $i$ grayscale value of the original image, $w$ and $h$ are the width and height of the image, $k$ is a number of color variations. The CLAHE has two variables controlling contrast image quality: the block size and clip limit [39]. The (4) is used to calculate the clip limit ($\beta$).

$$\beta = \frac{M}{N}\left(1 + \frac{\alpha}{100}(S_{max} - 1)\right) \quad (4)$$

Here, $N$ and $M$ denote the total number of grey-level pixels within each block while $S_{max}$ represents the highest permissible slope in the histogram's cumulative distribution function (CDF). This constraint helps to minimize artifacts by controlling the noise amplification. In addition, $\alpha$ is the clip factor, which ranges from 0 to 100.

## 2.4. Convolutional neural network

A CNN is a deep neural network that processes data such as images. Inspired by the visual cortex of animals, CNN has become the cornerstone of modern computer vision applications [40]. The architecture of a typical CNN consists of several layers, each serving a distinct purpose in the data processing pipeline. The primary layers included the following:
− The input layer accepts raw data, which are represented as multidimensional arrays of pixel values.
− In the convolutional layer, a series of learnable kernel filters perform convolution operations on the input data. These filters move across the input and perform element-wise multiplication and summation to generate the feature maps. The process was calculated using (5).

$$(I * K)(i,j) = \sum_m \sum_n I(i-m, j-n) \cdot K(m,n) \quad (5)$$

Where $I$ is the input image, $K$ is the kernel and $(i, j)$ are the coordinates of the output feature map.

- Activation layer: following each convolutional layer, an activation function is applied to introduce nonlinearity into the model. A rectified linear unit (ReLU) or Leaky ReLU was used in the activation layer. In (6) is used for the ReLU activation layer.

$$ReLU(x) = max(0, x) \tag{6}$$

- The spatial dimensions of the feature maps are reduced by the pooling layer, which helps lower the computational requirements and combat overfitting. Max pooling, illustrated in (7) and average pooling, as shown in (8), are the two most frequently employed pooling techniques.

$$(X)_{i,j,k} = \max_{m,n} X_{i.s_x+m.j.s_y+n,k} \tag{7}$$

$$(X)_{i,j,k} = \frac{1}{f_x \cdot f_y} \sum_{m,n} X_{i.s_x x+m.j.s_y+n,k} \tag{8}$$

- Fully connected layer: following multiple convolutional and pooling layers, the network performs advanced reasoning through fully connected layers. In these layers, each neuron is linked to every neuron in the preceding layer, enabling sophisticated input data representation.
- The output layer is the final layer of the CNN that outputs the predictions. The SoftMax activation function produces probability distributions for the target classes.

$$Softmax(x_i) = \frac{e^{x_i}}{\sum_{j=1}^{K} e^{x_j}} \tag{9}$$

## 3. RESULTS AND DISCUSSION

This study explored the impact of various image preprocessing techniques, including a PMD filter, CLAHE, and hybrid PMD filter-CLAHE. The evaluation used multiple performance metrics, including the CEIQ score for image quality assessment, accuracy, precision, recall, F1-score, and training time, to evaluate the preprocessing techniques using CNN models.

### 3.1. Images contrast assessment

The average CEIQ metric was evaluated on 10 pap smear images to assess the performance of different preprocessing techniques in enhancing pap smear image quality. Figure 3 shows the comparative results of these techniques (the first simulation). The original images achieved an average CEIQ of 3.5228. However, applying the PMD filter slightly reduced the score to 3.4956, suggesting that the PMD filter alone may not significantly enhance image contrast. However, CLAHE demonstrated a notable improvement, achieving a CEIQ score of 3.7301, highlighting its effectiveness in improving the contrast of cervical cell images. The hybrid approach, which combined the PMD filter with CLAHE, resulted in a CEIQ score of 3.7188, which was slightly lower than that of CLAHE alone but still substantially higher than that of the original and PMD filter images. These findings suggest that CLAHE plays a crucial role in contrast enhancement, whereas the hybrid PMD filter-CLAHE method effectively reduces noise while improving contrast. However, additional preprocessing with the PMD filter did not significantly enhance the contrast beyond that achieved using CLAHE alone. Overall, CLAHE was the most effective technique for contrast enhancement in the pap smear image analysis. In contrast, the hybrid PMD filter-CLAHE approach offers additional noise reduction and contrast improvement advantages.

### 3.2. CNN architectures assessment
#### 3.2.1. Pap smear images without preprocessing

The second simulation, shown in Table 1, presents the evaluation results of various CNN architectures on pap smear images without preprocessing. Among the tested models, EfficientNet-B0 achieved the highest accuracy (91.58%) and best F1-score (91.16%), demonstrating its superior ability to extract relevant features from cervical cell images. However, its training time (880 s) is relatively high compared to models such as MobileNet-V2 and SqueezeNet-1.0. EfficientNet-B1 closely followed, with an accuracy of 89.60% and an F1-score of 89.84%; however, it had the longest training time (1,049 s) among the EfficientNet models. MobileNet-V2 also delivered strong performance, with 83.91% accuracy and a relatively low training time (682 s), making it a good choice for balancing accuracy and computational efficiency. ResNet34 and DenseNet121 exhibited moderate performance, with an accuracy of 76.73%, whereas ResNet50 performed slightly worse at 71.53% accuracy. Although DenseNet201 achieved a slightly better accuracy (76.98%), it had a significantly longer training time (9,819 s). DenseNet201 is inefficient for large-scale implementations. SqueezeNet-1.0, designed for lightweight applications, achieves 73.76% accuracy with the fastest training time (632 s), but its lower F1-score (72.56%) suggests a trade-off between speed and performance.

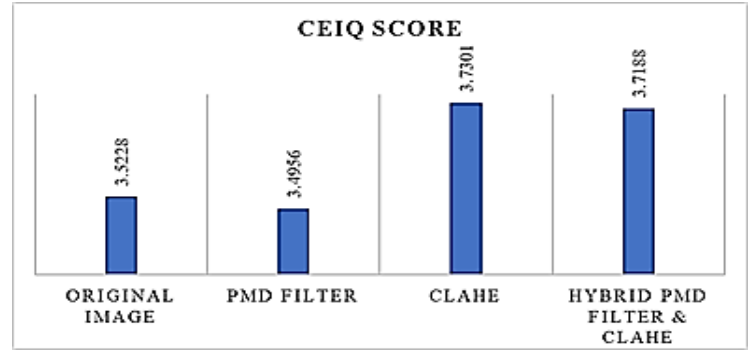

Figure 3. Average CEIQ score for image enhancement

Table 1. CNN Architectures' assessment score for pap smear images without preprocessing

| Architecture | Accuracy | Precision | Recall | F1-score | Training time (s) |
|---|---|---|---|---|---|
| ResNet34 | 76.73% | 79.12% | 76.90% | 76.19% | 755 |
| ResNet50 | 71.53% | 75.66% | 71.64% | 70.48% | 892 |
| SqueezeNet-1.0 | 73.76% | 73.99% | 74.07% | 72.56% | 632 |
| MobileNet-V2 | 83.91% | 85.71% | 84.10% | 83.63% | 682 |
| EfficientNet-B0 | **91.58%** | **92.30%** | **91.68%** | **91.16%** | 880 |
| EfficientNet-B1 | 89.60% | 91.55% | 89.64% | 89.84% | 1,049 |
| DenseNet121 | 76.73% | 79.81% | 76.85% | 76.04% | 964 |
| DenseNet201 | 76.98% | 80.02% | 77.16% | 75.93% | 9,819 |

Overall, EfficientNet-B0 emerged as the best-performing model regarding accuracy, precision, recall, and the F1-score. EfficientNet-B0 was the most suitable choice for cervical cancer classification in this simulation. However, if computational efficiency is a priority, MobileNet-V2 provides a strong alternative with relatively high accuracy and significantly lower training time. The results suggest that deeper architectures, such as EfficientNet and MobileNet-V2, are more effective for feature extraction. In contrast, lighter models, such as SqueezeNet and ResNet, require further optimization to achieve competitive performance.

### 3.2.2. Pap smear images with PMD filter preprocessing

Table 2 presents the performance evaluation results of the CNN architecture on the pap smear images preprocessed with the PMD filter technique in the third simulation. EfficientNet-B0 remained the best-performing model, achieving an accuracy of 88.86%, slightly lower than its performance on the original images (91.58%). Similarly, EfficientNet-B1 experienced a drop in accuracy from 89.60% to 85.89%, suggesting that, although the PMD filter may enhance noise reduction, it does not necessarily improve the classification performance of these architectures. Interestingly, SqueezeNet-1.0 and ResNet34 exhibited notable improvements with the PMD filter. SqueezeNet-1.0 improves its accuracy from 73.76% to 80.20%, demonstrating that the PMD filter enhances feature extraction for lightweight architectures. The accuracy of ResNet34 also increased from 76.73% to 79.21%, indicating that noise reduction contributed to better classification results in this architecture. The accuracy of ResNet50 improved from 71.53% to 77.23%, although it remained less competitive than those of the other models.

On the other hand, MobileNet-V2, which performed well on the original images (83.91% accuracy), decreased slightly to 81.93% with the PMD filter. A similar decline was observed in DenseNet121 (76.73% to 78.22%) and DenseNet201 (76.98% to 75.50%), with DenseNet201 experiencing the most significant drop. This suggests that deeper architectures that rely on high-detail feature extraction may not benefit as much from the PMD filter, potentially because the smoothing effect reduces finer details in cervical cell structures. From this comparison, lightweight architectures such as SqueezeNet-1.0, ResNet34, and ResNet50 benefit the most from the PMD filter, whereas deeper networks such as EfficientNet-B0, EfficientNet-B1, and DenseNet201 show a decline in their performance. This suggests that the PMD filter is particularly effective for models that struggle with noise; however, it may remove fine-grained details on which deeper architectures rely on classification.

Table 2. CNN Architectures' assessment score for pap smear images using PMD filter preprocessing

| Architecture | Accuracy | Precision | Recall | F1-score | Training time (s) |
|---|---|---|---|---|---|
| ResNet34 | **79.21%** | **80.49%** | **79.39%** | **78.76%** | 784 |
| ResNet50 | **77.23%** | **77.76%** | **77.51%** | **76.60%** | 897 |
| SqueezeNet-1.0 | **80.20%** | **79.33%** | **80.38%** | **79.29%** | 633 |
| MobileNet-V2 | 81.93% | 84.63% | 82.07% | 81.42% | 685 |
| EfficientNet-B0 | 88.86% | 90.32% | 88.99% | 88.19% | 858 |
| EfficientNet-B1 | 85.89% | 88.66% | 85.89% | 85.56% | 1,048 |
| DenseNet121 | **78.22%** | **82.73%** | **78.47%** | **77.42%** | 1,043 |
| DenseNet201 | 75.50% | 77.96% | 75.68% | 74.19% | 11,196 |

### 3.2.3. Pap smear images with CLAHE preprocessing

Table 3 presents the evaluation results of the CNN architectures on the pap smear images preprocessed using the CLAHE technique in the fourth simulation. The evaluation of CNN architectures on CLAHE preprocessed pap smear images demonstrated significant improvements in classification performance compared to the original and PMD-filtered images. CLAHE enhances contrast, crucial for improving feature extraction in cervical cell classification. Among all the architectures, EfficientNet-B1 achieved the highest accuracy (89.36%), followed by EfficientNet-B0 (88.86%) and MobileNet-V2 (84.65%). Interestingly, ResNet34 showed a substantial improvement, reaching an accuracy of 84.16% compared with 79.21% with the PMD filter and 76.73% with the original images. Similarly, ResNet50 improved to 82.92%, indicating a better adaptation to contrast-enhanced images. DenseNet121 and DenseNet201 also exhibited considerable improvements, with DenseNet121 increasing from 78.22% (PMD) to 83.17% (CLAHE) and DenseNet201 increasing from 75.50% (PMD) to 80.69% (CLAHE). This suggests that CLAHE enhances the performance of deeper architectures by improving contrast, allowing them to extract more meaningful features.

However, SqueezeNet-1.0, which showed a notable improvement with the PMD filter (80.20%), did not benefit as much from CLAHE and achieved an accuracy of 74.75%. This suggests that lightweight models may be more sensitive to noise reduction (PMD) than contrast enhancement (CLAHE). CLAHE enhanced the classification performance across most architectures, particularly deeper CNNs, such as ResNet34, ResNet50, DenseNet121, and DenseNet201. EfficientNet-B0 and EfficientNet-B1 remained strong performers; however, their best accuracy was still observed on the original images rather than with PMD or CLAHE preprocessing. SqueezeNet-1.0 benefits more from the PMD filter than CLAHE, suggesting that noise reduction is more important for this lightweight model than contrast enhancement. MobileNet-V2 achieved its highest accuracy (84.65%) with CLAHE, slightly outperforming the original and the PMD filter versions.

Table 3. CCN Architectures' assessment score for pap smear images using CLAHE preprocessing

| Architecture | Accuracy | Precision | Recall | F1-score | Training time (s) |
|---|---|---|---|---|---|
| ResNet34 | **84.16%** | **86.75%** | **84.40%** | **84.02%** | 790 |
| ResNet50 | **82.92%** | **83.48%** | **83.06%** | **82.57%** | 905 |
| SqueezeNet-1.0 | **74.75%** | **76.33%** | **75.01%** | **74.41%** | 631 |
| MobileNet-V2 | **84.65%** | **85.18%** | **84.85%** | **84.17%** | 687 |
| EfficientNet-B0 | 88.86% | 89.18% | 88.96% | 88.19% | 842 |
| EfficientNet-B1 | 89.36% | 90.35% | 89.47% | 89.11% | 1,053 |
| DenseNet121 | **83.17%** | **84.85%** | **83.40%** | **82.56%** | 972 |
| DenseNet201 | **80.69%** | **83.64%** | **80.92%** | **79.67%** | 9,880 |

### 3.2.4. Pap smear images with hybrid PMD filter-CLAHE preprocessing

Table 4 presents the evaluation results of the CNN architectures on the pap smear images preprocessed using the hybrid PMD filter-CLAHE technique in the final simulation. The hybrid approach combines noise reduction from the PMD filter with contrast enhancement from CLAHE, which enhances the ability of the CNN models to extract meaningful features from images. Among all architectures, EfficientNet-B0 achieved the highest accuracy (91.74%), slightly surpassing its previous performance with the original images (91.58%) and CLAHE (88.86%). This suggests that combining noise reduction and contrast enhancement benefits this architecture more than CLAHE or PMD alone. Similarly, EfficientNet-B1 maintained a high accuracy (89.77%), comparable to its original image performance (89.60%) and slightly better than CLAHE (89.36%). DenseNet121 and DenseNet201 showed notable improvements, with DenseNet121 increasing to 86.39% (from 83.17% with CLAHE to 78.22% with PMD), whereas DenseNet201 reached 83.66%, surpassing CLAHE (80.69%) and PMD (75.50%) performance. This indicates that the DenseNet models benefit significantly from noise reduction and contrast enhancement. The ResNet architectures also performed well, with ResNet50 achieving 85.15% accuracy and improving both PMD (77.23%) and CLAHE (82.92%). ResNet34 reaches 84.41%, showing a similar trend. This confirmed that contrast enhancement with CLAHE was adequate; however, noise reduction further refined the extracted features, leading to better classification accuracy.

However, SqueezeNet-1.0 did not benefit as much from the hybrid approach, achieving an accuracy of 78.47%, which is lower than its PMD-only result (80.20%), but higher than its CLAHE result (74.75%). This suggests that lightweight architectures like SqueezeNet may not gain significant advantages from extensive preprocessing and may be more optimized for straightforward noise reduction techniques. The hybrid PMD FILTER-CLAHE approach achieved the best overall performance for most CNN architectures, particularly for deeper networks such as EfficientNet-B0, EfficientNet-B1, DenseNet121, and DenseNet201. ResNet architectures benefit significantly from the hybrid approach, with ResNet50 improving the most (from 71.53% in the original to 85.15% with hybrid processing). SqueezeNet-1.0 performs best with the PMD filter alone (80.20%), indicating that lightweight architectures may benefit more from noise reduction than contrast enhancement. EfficientNet-B0 achieved the highest accuracy (91.74%) with hybrid processing, slightly outperforming the original image accuracy (91.58%). The hybrid PMD filter-CLAHE technique provides an optimal balance for preprocessing pap smear images, maximizing the classification performance across multiple CNN architectures. This was the most effective preprocessing method used in this study.

Table 4. CNN Architectures' assessment score for pap smear images using a hybrid PMD filter-CLAHE preprocessing technique

| Architecture | Accuracy | Precision | Recall | F1-score | Training time (s) |
|---|---|---|---|---|---|
| ResNet34 | **84.41%** | **85.78%** | **84.57%** | **83.09%** | 749 |
| ResNet50 | **85.15%** | **85.70%** | **85.32%** | **84.82%** | 894 |
| SqueezeNet-1.0 | **78.47%** | **81.85%** | **78.71%** | **77.24%** | 630 |
| MobileNet-V2 | **83.91%** | **86.56%** | **84.18%** | **83.25%** | 685 |
| EfficientNet-B0 | **91.74%** | 91.52% | 91.28% | **91.33%** | 862 |
| EfficientNet-B1 | **89.77%** | 88.75% | 88.41% | **89.94%** | 1,052 |
| DenseNet121 | **86.39%** | **87.47%** | **86.57%** | **86.21%** | 955 |
| DenseNet201 | **83.66%** | **85.29%** | **83.91%** | **83.13%** | 10,305 |

## 4. CONCLUSION

Cervical cancer is a major health concern, especially in developing countries, where early detection is crucial. Deep learning, particularly CNNs, has shown promise for automated cervical cancer classification, but pap smear image quality significantly affects performance. This study evaluated the effects of the PMD filter, CLAHE, and their hybrid approach on cervical cancer classification. Results showed that the hybrid PMD filter-CLAHE outperformed individual preprocessing methods, with maximum improvements of 13.62% in accuracy, 10.04% in precision, 13.08% in recall, and 14.34% in F1-score. PMD filtering was particularly effective in lightweight CNN architectures, while CLAHE benefited deeper architectures. The hybrid approach balanced noise reduction and contrast enhancement, improving performance across most CNNs without significantly increasing processing time. Although this study demonstrated the benefits of hybrid preprocessing, further research is needed to confirm its effectiveness across different imaging conditions, staining techniques, and datasets. Larger-scale validation and real-time implementation in clinical workflows could enhance its practical application. These findings highlight the potential of combining noise reduction and contrast enhancement to improve CNN-based cervical cancer classification. Future research should focus on optimizing preprocessing parameters using advanced algorithms and integrating real-time methods to enhance accuracy and efficiency.

**FUNDING INFORMATION**

This research was funded by the Ministry of Higher Education, Science, and Technology of the Republic of Indonesia and the Indonesian Education Foundation (LPDP) through the Center for Higher Education Funding and Assessment (PPAPT) under the Indonesian Education Scholarship (BPI) program.

AUTHOR CONTRIBUTIONS STATEMENT
All authors contributed as follows and have read and approved the final manuscript.

| Name of Author | C | M | So | Va | Fo | I | R | D | O | E | Vi | Su | P | Fu |
|---|---|---|---|---|---|---|---|---|---|---|---|---|---|---|
| Ach Khozaimi | ✓ | ✓ | ✓ | ✓ | ✓ | ✓ |  | ✓ | ✓ |  | ✓ |  | ✓ | ✓ |
| Isnani Darti |  | ✓ |  | ✓ | ✓ |  | ✓ |  |  | ✓ |  | ✓ | ✓ |  |
| Syaiful Anam | ✓ | ✓ | ✓ |  |  | ✓ | ✓ | ✓ |  | ✓ | ✓ | ✓ |  |  |
| Wuryansari Muharini Kusumawinahyu | ✓ | ✓ |  | ✓ |  |  |  |  |  | ✓ |  | ✓ | ✓ |  |

| | | | | | |
|---|---|---|---|---|---|
| C | : | **C**onceptualization | I | : | **I**nvestigation |
| M | : | **M**ethodology | R | : | **R**esources |
| So | : | **So**ftware | D | : | **D**ata Curation |
| Va | : | **Va**lidation | O | : | Writing - **O**riginal Draft |
| Fo | : | **Fo**rmal analysis | E | : | Writing - Review & **E**diting |

| | | |
|---|---|---|
| Vi | : | **Vi**sualization |
| Su | : | **Su**pervision |
| P | : | **P**roject administration |
| Fu | : | **Fu**nding acquisition |

CONFLICT OF INTEREST STATEMENT
The authors confirm that they have no conflicts of interest regarding the publication of this work and no financial or personal affiliations that could affect its content.

DATA AVAILABILITY
The publicly available SIPaKMeD dataset was used in this study and can be accessed at https://bit.ly/SIPaKMeD for further research in medical image processing and cervical cancer classification.

## BIOGRAPHIES OF AUTHORS

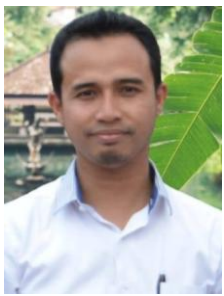
**Ach Khozaimi** is a lecturer in informatics engineering at Trunojoyo University of Madura, Indonesia. He earned his bachelor's degree in informatics from Trunojoyo University of Madura and his master's degree in informatics from Institut Teknologi Sepuluh Nopember (ITS), Indonesia. Currently, he is pursuing a Ph.D. in the Mathematics Department at Brawijaya University, Indonesia. His doctoral studies are funded by the Center for Higher Education Funding and Assessment (PPAPT) under the Ministry of Higher Education, Science, and Technology of the Republic of Indonesia. He can be contacted at email: khozaimi@trunojoyo.ac.id.

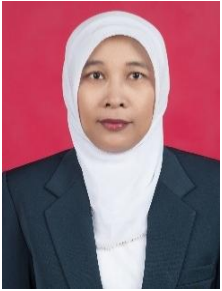
**Isnani Darti** 🆔 📧 SC 📇 stands as the 24th active Professor at the Faculty of Mathematics and Natural Sciences (FMIPA). Her research interests span several captivating domains: applied dynamical systems, mathematical biology, optical solitons, and discretization of continuous dynamical systems. Recently, she achieved the prestigious rank of Professor at Brawijaya University. She is the 24th active Professor at the Faculty of Mathematics and Natural Sciences (FMIPA) and the 176th active at the university overall. Her professorship adds to the rich legacy of Brawijaya University, where she is the 335th Professor in its history. She can be contacted at email: isnanidarti@ub.ac.id.

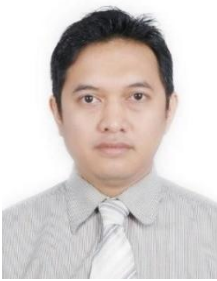
**Syaiful Anam** 🆔 📧 SC 📇 received a Doctor of Natural Science and Mathematics degree from Yamaguchi University, Japan in 2015. He also received his Bachelor's Degree in Mathematics from Brawijaya University, Indonesia in 2001 and his Master Degree from Sepuluh Nopember Institute of Technology, Indonesia in 2006. He is currently an assistant professor at Mathematics Department, Brawijaya University, Malang, Indonesia. His research includes data science, computational intelligence, machine learning, digital image processing, and computer vision. He has published over 35 papers in international journals and conferences. He can be contacted at email: syaiful@ub.ac.id.

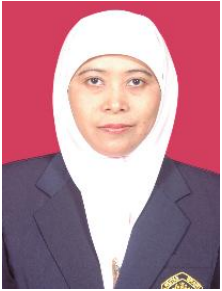
**Wuryansari Muharini Kusumawinahyu** 🆔 📧 SC 📇 a lecturer at the Department of Mathematics, Faculty of Science, Universitas Brawijaya in Malang, Indonesia. She is actively involved in teaching and research. Here's a summary of his educational background: Bachelor's Degree (S1), Master's Degree (S2) and mathematics, and Doctoral Degree (S3) from Institute Teknologi Bandung, Bandung, Indonesia. Her research interests span various mathematical topics, and she has contributed to several areas, including epidemiology, predator-prey models, and wave dynamics. She can be contacted at email: wmuharini@ub.ac.id.